\documentclass[journal,twoside,web]{ieeecolor}
\usepackage{lcsys}
\usepackage{verbatim,xpatch}
\usepackage[export]{adjustbox}
\usepackage{amsmath,amssymb}
\usepackage[scaled]{helvet}
\usepackage[T1]{fontenc}

\usepackage{etoolbox}
\makeatletter
\@ifundefined{color@begingroup}%
  {\let\color@begingroup\relax
   \let\color@endgroup\relax}{}%
\def\fix@ieeecolor@hbox#1{%
  \hbox{\color@begingroup#1\color@endgroup}}
\patchcmd\@makecaption{\hbox}{\fix@ieeecolor@hbox}{}{\FAILED}
\patchcmd\@makecaption{\hbox}{\fix@ieeecolor@hbox}{}{\FAILED}

\def\BibTeX{{\rm B\kern-.05em{\sc i\kern-.025em b}\kern-.08em
    T\kern-.1667em\lower.7ex\hbox{E}\kern-.125emX}}
\usepackage{graphicx, subfig, booktabs}
\usepackage{amsmath,amssymb,amsthm, bm}
\usepackage{cases}
\usepackage[T1]{fontenc}
\usepackage{hyperref}
\usepackage{newtxmath}
\allowdisplaybreaks
\hypersetup{
     colorlinks =true,
     linkcolor = blue,
     filecolor = blue,
     citecolor = magenta,      
     urlcolor = magenta,
     }
\usepackage{setspace}
\usepackage{siunitx}
\usepackage{csvsimple}
\usepackage{xurl}
\usepackage{cite}
\usepackage{algorithm}
\usepackage{algpseudocode}

\let\labelindent\relax
\usepackage{enumitem, xspace}

\usepackage{accents}
\newcommand{\ie}{i.e.}
\renewcommand{\bf}[1]{\mathbf{#1}}
\newcommand{\bs}[1]{\boldsymbol{#1}}

\newcommand{\mR}{\mathbb{R}}

\theoremstyle{remark}
\newtheorem{remark}{Remark}
\newtheorem{definition}{Definition}

\newtheorem{lemma}{Lemma}
\newtheorem{theorem}{Theorem}

\begin{document}


\title{ The Solution of Potential-Driven, Steady-State Nonlinear Network Flow Equations via Graph Partitioning}

\author{Shriram Srinivasan$^{\dagger}$, Kaarthik Sundar$^{\dagger}$
\thanks{$^{\dagger}$ Los Alamos National
Laboratory, Los Alamos, New Mexico, USA. E-mail: \texttt{\{shrirams,kaarthik\}@lanl.gov}}
}

\maketitle
\thispagestyle{empty}

\begin{abstract}
The solution of potential-driven steady-state flow in large networks is required in various engineering applications, such as transport of natural gas or water through pipeline networks. The resultant system of nonlinear equations depends on the network topology, and its solution grows more challenging as the network size increases.
We present an algorithm  that utilizes a given partition of a
network into tractable sizes to compute
a global solution for the full nonlinear system through local
solution of smaller subsystems induced by the partitions. 
When the partitions are induced by interconnects or transfer points corresponding to networks owned by different operators, the method ensures data is shared solely at the interconnects, leaving  network operators free to  solve the network flow system corresponding to  their own domain in any manner of their choosing. 
The proposed method is shown to be connected to the Schur complement and the  method's viability demonstrated on some challenging test cases.
%
\end{abstract}

\IEEEpeerreviewmaketitle

\def\d{\partial}

\section{Introduction}
\label{sec:intro}

In the control of networked flow systems, one of the most fundamental  and frequently performed tasks is the simulation of steady-state network flow under given control actions. 
When control parameters produce states that stray into regimes far from normal operating conditions, the subsequent contingency analysis must be done over as large a scale as possible to be useful.
However, the gas pipeline network  in the continental US viewed on a national scale is a patchwork of multiple regional and local networks,  owned and operated individually but held together as a single connected network  through interconnects or transfer points. 
A simulation of network flow contingencies over the national scale is difficult due to two disparate reasons. 

{On one hand, there are concerns of privacy and issues with data sharing, which make it difficult to perform simulations and study contingency scenarios over regions of the US pipeline network where multiple operators are involved. 
Privacy concerns prevent the assembly, storage, and formation of the monolithic system corresponding to a large composite network composed of the interconnected union of multiple independent networks. A viable algorithm needs to solve the monolithic system while respecting the data-sharing constraints imposed on it by exchanging information only at the interconnects or transfer points between networks.
On the other hand, the mathematical problem is difficult due to the sheer size and scale of the network. 
The standard Newton-Raphson iterative scheme for the nonlinear system faces challenges as the network size increases. Even the solution of the linear system in each Newton step becomes expensive with direct methods like LU decomposition, which scale as $O(N^3)$ with system size and are memory-intensive. However, indirect or iterative methods based on Krylov subspace solvers seem to perform poorly \cite{Nguyen2025Jun} due to ill-conditioning.}

{A remedy for both these ills would be to partition the network into tractable sizes (so that the condition number is reduced) and devise a method to compute a global solution for the full nonlinear system through local solution of smaller subsystems induced by the partitions.}
There have only been a couple of relevant efforts in this direction. An early attempt \cite{Rios-Mercado2002Nov} had unsubstantiated assumptions and was narrow in scope. More recently, a solution technique was devised based on a hierarchical partitioning of the network \cite{network-partitioning} that utilizes the presence of special junctions called \emph{articulation points}. 
While useful in its own right, the hierarchical solution technique proposed in \cite{network-partitioning}  is still limited in its ability to handle the interconnects or transfer points at the interface between multiple regional and local gas networks, since these interconnects are typically not articulation points.

The more general solution method proposed in this article relates to the Schur complement, which insinuates itself into well-known techniques known in different areas of research. 
Depending on the reader's persuasion \cite{levitt}, our method could be said to share similarities with domain decomposition methods and the Dirichlet-to-Neumann map for the solution of PDEs,  or the Kron reduction for circuits \cite{vanderSchaft2024May}. 
However, it differs from both in there being no explicit elimination of any degrees of freedom. 
The method proposed allows for data sharing solely at the interconnects or transfer points between individual subnetworks, with network operators free to use their own solvers or software to solve the network flow system for their own domain. 
In that context, we present a method that utilizes a general partition of the network, is in the spirit of the suggested remedy, and avoids the limitations of the solution technique \cite{network-partitioning} that was based on hierarchical partitioning.

\section{Governing equations}
\label{sec:model-equations}
We stick to the same notation as in \cite{network-partitioning} to describe the potential-driven steady-state flow equations on a network. Even though the equations and concepts are identical, it was necessary to include them for completeness.

Let $G$ denote the graph of the network, where $V(G)$ and $E(G)$ are the corresponding sets of \emph{nodes} and \emph{edges} of the network.   
For each node $i \in V(G)$, we let $\pi_i \in \mR$ denote the nodal value of the potential while $q_i \in \mR$ is the nodal injection or supply.  
We let $f_{ij} \in \mR$ denote the steady flow through the edge  $(i, j) \in E(G)$  that connects nodes $i, j \in V(G)$. 
In order to ensure the uniqueness of the potentials satisfying the equations, it is sufficient (but not always necessary, see \cite{network-partitioning} for details) that the potential be known or fixed at certain nodes.
Hence, each node $i \in V(G)$  is assigned to one of two mutually disjoint subsets of $V(G)$ consisting of so-called slack nodes $N_s$ and non-slack nodes $N_{ns}$ respectively, depending on whether the potential $\pi_i^*$ or injection $q_i^*$ at the node is specified. 
Thus $V(G) = N_s \cup N_{ns}$ and one writes the system of network flow equations consisting of balance of flows at each node, potential-driven flow on each edge, and the specified potential at the slack nodes  in the general form ($*$ superscript indicates given data)
\begin{subnumcases} {\label{eq:NF-pi}
\mathcal{NF}(G):}  
\gamma^{*}_{ij}\pi_i - \pi_j = g^{*}_{ij}(f_{ij}) \; \forall (i, j) \in E(G),  \label{eq:NF-pi-edge}\\ 
\sum_{(i, j) \in E(G)} f_{ij}  -  \sum_{(j, i) \in E(G)} f_{ji} = q^{*}_i \; \forall i \in N_{ns}, \label{eq:NF-pi-balance}\\
\pi_i = \pi^{*}_i \; \forall i \in N_s. \label{eq:NF-pi-slack}
\end{subnumcases}
%
%

%
\begin{remark}
\label{rem:slack-injection}
Note that if $|N_s| > 1$, then $\forall i \in N_s$, $q_i$ is unknown and can be determined only once the edge flows $f_{ij}$ are known for all $(i, j) \in E(G)$, \ie, after solving the system  \eqref{eq:NF-pi}.
However, if $|N_s| =1$, the steady state condition that the sum of all the nodal injections must be zero is enough to determine the injection at the slack node.
\end{remark}
In this general form, $g^{*}_{ij}: \mathbb{R} \mapsto \mathbb{R}$ is a  monotonic function, $q^{*}_i \in \mathbb{R}$ while  $\gamma^{*}_{ij} \in \mathbb{R}^{+}$.
In network flow applications, different types of active or passive edge elements can be modelled through various choices and specifications of $g^{*}_{ij}$ and $\gamma^{*}_{ij}$. 
In order to model the flow of fluid in pipeline networks, the pipes are modelled as edge elements by setting  $\gamma^{*}_{ij} =1$ and the nonlinear resistance function $g^{*}_{ij}(f_{ij}) \propto f_{ij} |f_{ij}|$ \cite{nrsolver}. 
Additionally, for compressible fluids such as natural gas or hydrogen \cite{kazi2024modeling}, the choice $g^{*}_{ij} = 0$ corresponds to pressure regulators or compressors depending on whether $\gamma^{*}_{ij}$ is less than or equal to unity \cite{Schmidt2017}.
In contrast,  for incompressible liquids such as water, pumps are modelled as edges where 
$\gamma^{*}_{ij} =1$ and $g^{*}_{ij}$ is some positive constant or a monotonic function of $f_{ij}$ \cite{singh2019optimal}. 

When certain assumptions on the network topology are valid, the solution to \eqref{eq:NF-pi}  is known to be unique (when it exists).
However, these assumptions also eliminate certain impractical but pathological network instances \cite{nrsolver}, and are stated as follows :
\begin{enumerate}[label=(A\arabic*)]
    \item There is at least one slack node, i.e., $|N_s| \geqslant 1$.\label{assumption:slacks}
    \item When  $|N_s| \geqslant 2$, a path connecting two slack nodes must consist of at least one edge with $g^{*}_{ij} \neq 0$ \label{assumption:path-pipe}
    \item Any cycle must consist of at least one edge with $g^{*}_{ij} \neq 0$. \label{assumption:cycle-pipe}
\end{enumerate}
We already mentioned that \ref{assumption:slacks} ensures that the values of the potential satisfying equation \eqref{eq:NF-pi} are unique, but \ref{assumption:path-pipe} and \ref{assumption:cycle-pipe} are required  to ensure uniqueness of the flows satisfying \eqref{eq:NF-pi} \cite{nrsolver, Singh2019, Singh2020}.
The existence of a solution to \eqref{eq:NF-pi} is not discussed in this article since the proof of existence of a solution  for a particular case of \eqref{eq:NF-pi} can be found in \cite{ss-soln-existence}.

For any \emph{nonlinear} edge relation, including the aforementioned case of fluid flow in pipeline networks,  \eqref{eq:NF-pi} is a nonlinear system of algebraic equations in $|V(G)| + |E(G)|$ variables whose solution becomes more challenging as the network size grows larger.  
Thus, there is a strong motivation to develop a solution technique that can utilize network partitioning to solve the system \eqref{eq:NF-pi} on smaller subnetworks.
While \cite{network-partitioning} proposed such a method, it has severe limitations imposed by the network structure and the placement of slack nodes (which will be discussed subsequently), but the method we propose overcomes most of those deficiencies.

As a preamble to it, we present a reformulation of the system ~\eqref{eq:NF-pi} as a bi-level nonlinear system. The reformulation   will also yield useful insights for the subsequent network partitioning.

\section{A bi-level reformulation}
\label{sec:bi-level}
Let $\bm A$ be the edge incidence matrix of the graph $G$  of size $|V(G)| \times |E(G)|$. Each element of $\bm A$ is defined as follows: 
\begin{flalign}
A_{ij} = \begin{cases}
-1 & \text{ if $(i, j) \in E(G)$, } \\
+1 & \text{ if $(j, i) \in E(G)$, } \\
0 & \text{ otherwise. }
\end{cases} \label{eq:incidence}
\end{flalign}
Define a weighted edge incidence matrix  $\bm B$ as: 
\begin{flalign}
B_{ij} = \begin{cases}
-1 & \text{ if $(i, j) \in E(G)$, } \\
+\gamma^*_{ij} & \text{ if $(j, i) \in E(G)$, } \\
0 & \text{ otherwise. }
\end{cases} \label{eq:wt-incidence}
\end{flalign}
We already know that $V(G) = N_{ns} \cup N_s$. Let us divide the nonslack vertices in $N_{ns}$ into two disjoint, non-empty subsets, $N_{ns} = N_{\textrm{int}} \cup N_{\textrm{dep}}$, with the subscripts meant to stand for \emph{interface} nodes and \emph{dependent} nodes, respectively. Typically, $|N_{\textrm{int}}| \ll |N_{\textrm{dep}}|$.
The nodes can be ordered such that the following partitioning of $\bs{A}$ into \emph{dependent}, \emph{interface} and \emph{slack} makes sense:
\begin{equation}
    \label{eq:A-block}
    \bs{A} = \begin{bmatrix}  \bs{A}_{\textrm{dep}} \;\; \bs{A}_{\textrm{int}} \;\; \bs{A}_{s}    \end{bmatrix}^T
\end{equation}
An analogous partition of $\bm B$ will be assumed. In the same vein, we define 
\begin{flalign*} 
& \text{nodal potentials: } \bm \pi \in \mR^{|V(G)|} = \begin{bmatrix} \bm \pi_{\textrm{dep}} \;\; \bm \pi_{\textrm{int}} \;\; \bm \pi_s \end{bmatrix}^T, \\ 
& \text{edge flows: }  \bm f \in \mR^{|E(G)|}, \text{ and }\\ 
& \text{non-slack nodal injections: }   \bm q^* \in \mR^{|N_{ns}|} = \begin{bmatrix} \bm q^*_{\textrm{dep}} \;\; \bm q^*_{\textrm{int}} \end{bmatrix}^T.
\end{flalign*}
Finally, we let $\bm g^*(\bm f)$ the vector of edge-functions $g^*_{ij}(f_{ij})$ for each $(i,j) \in E(G)$.
In this notation, with the aid of \eqref{eq:A-block}, one can rewrite the system \eqref{eq:NF-pi} equivalently as
\begin{subequations} 
\begin{flalign}
h_{\mathrm{edge}}(\bm f, \bm \pi) \triangleq &~ \bm{B}^T  \bm \pi - \bm g^*(\bm f) = \bm 0,  \label{eq:NF-pi-edge-m}\\ 
h_{\mathrm{dep}}(\bm f) \triangleq &~ \bm{A}_{\textrm{dep}}  \bm f - \bm q^*_{\textrm{dep}} = \bm 0, \label{eq:NF-pi-balance-m-1}\\
h_{\mathrm{int}}(\bm f) \triangleq &~ \bm{A}_{\textrm{int}}  \bm f - \bm q^*_{\textrm{int}} = \bm 0, \label{eq:NF-pi-balance-m-2}\\
h_{\mathrm{slack}}(\bm \pi) \triangleq &~ \bm \pi_s - \bm \pi^{*} = \bm 0. \label{eq:NF-pi-slack-m}
\end{flalign}
\label{eq:NF-matrix}
\end{subequations}
In order to to express the flows $\bm f$ and the potentials $\bm \pi_{\textrm{dep}}$ in the system~\eqref{eq:NF-matrix} as functions of $\bm \pi_{\textrm{int}}$ and $\bm \pi_s$,  we define
\begin{flalign}
    \bm y \triangleq \begin{bmatrix} \bm f \\ \bm \pi_{\textrm{dep}} \end{bmatrix} \text{ and }
    H\left (\bm y , \bm \pi_{\textrm{int}}, \bm \pi_s \right) \triangleq \begin{bmatrix}
        h_{\mathrm{edge}}(\bm f, \bm \pi) \\ h_{\mathrm{dep}}(\bm f) \end{bmatrix}. \label{eq:setup-ift}
\end{flalign}
If we assume the Jacobian $D_{\bm y} H$ is non-singular (this holds under standard regularity and non-zero  flows \cite{nrsolver}),  then for a small enough neighbourhood around $\bm \pi_{\textrm{int}}$ and $\bm \pi_s$ , the implicit function theorem asserts the existence of a smooth mapping $\hat{\bm y}$ such that
\begin{subequations}
\begin{gather}
    \bm y = \hat{\bm y}(\bm \pi_{\textrm{int}}, \bm \pi_s), \\
    H( \hat{\bm y}(\bm \pi_{\textrm{int}}, \bm \pi_s), \bm \pi_{\textrm{int}}, \bm \pi_s) = \bm 0, \\ 
    D_{\bm \pi_{\textrm{int}}}\hat{\bm y}  = - \left[D_{\bm y} H\right]^{-1}D_{\bm \pi_{\textrm{int}}} H. \label{eq:ift-derivative}
\end{gather}
\label{eq:ift}
\end{subequations}
%
%
\begin{remark}
\label{rem:sensitivity}
    It will be useful to regard the formula for the derivative in equation \eqref{eq:ift-derivative}  as expressing the \emph{sensitivities} of the solution $\hat{\bm y}$ with respect to the parameters $\bm \pi_{\textrm{int}}$. 
    Thus the computation of the sensitivity is straightforward (solution of a linear system)  once the solution of the nonlinear system $H\left (\bm y , \bm \pi_{\textrm{int}}, \bm \pi_s \right)=\bm 0$ is known.
\end{remark}
Let us define an appropriate projection $\mathbb{P} = \begin{bmatrix}  \bm I & \bm 0\end{bmatrix}$ such that
$$\mathbb{P}\bm y = \bm f.$$
Then, using \eqref{eq:ift} and the equation \eqref{eq:NF-pi-slack-m}, we can rewrite \eqref{eq:NF-pi-balance-m-2} as
\begin{equation}
\label{eq:outer}
   h(\bm \pi_{\textrm{int}}) \triangleq h_{\mathrm{int}}(\mathbb{P}\bm y) = \bm{A}_{\textrm{int}}\; \mathbb{P}  \hat{\bm y}(\bm \pi_{\textrm{int}}, \bm \pi^*) - \bm q^*_{\textrm{int}} = \bm 0
\end{equation}
The inference  is that we can  solve \eqref{eq:outer} (a small system with $|N_{\textrm{int}}|$ variables), in order to find a solution to \eqref{eq:NF-matrix}.
However, a single step of the Newton-Raphson algorithm applied to \eqref{eq:outer} is
\begin{equation}
    \left[D_{\bm \pi_{\textrm{int}}}h \right]\Delta \bm \pi_{\textrm{int}} = -h(\bm \pi_{\textrm{int}}),
    \label{eq:NR-outer}
\end{equation}
where the Jacobian $D_{\bm \pi_{\textrm{int}}}h$ on using \eqref{eq:ift-derivative} is
\begin{subequations}
\begin{gather}
     D_{\bm \pi_{\textrm{int}}}h = - \bm{A}_{\textrm{int}}\; \mathbb{P} \; \left[D_{\bm y} H\right]^{-1}\left[ D_{\bm \pi_{\textrm{int}}} H \right],  \label{eq:jac-outer-1}\\
    \text{where, } H\left (\bm y , \bm \pi_{\textrm{int}}, \bm \pi^* \right) = \bm 0. \label{eq:jac-outer-2}
\end{gather}
\label{eq:Jacobian-outer}
\end{subequations}
Hence, every Newton iteration in the solution of \eqref{eq:outer} involves the solution of $H\left (\bm y , \bm \pi_{\textrm{int}}, \bm \pi^* \right) = \bm 0$ justifying the 
\begin{definition}[Bi-level formulation]
\label{def:bi-level}
 The vector $\bm \pi_{\textrm{int}}$ that solves  \eqref{eq:outer} through a sequence of Newton steps \eqref{eq:NR-outer} is a bi-level formulation of \eqref{eq:NF-matrix}, as depicted in Algorithm~\ref{alg:bilevel-NR}.
\end{definition}
\begin{algorithm}[H]
\caption{Newton Raphson for the bi-level formulation}
\begin{algorithmic}[1]
\State \textbf{input:} $\bm{\pi}_{\textrm{int}}^{(0)}$, $i \gets 0$\Comment{Initialization}
\Repeat
\State \label{step:solve-H}Solve $H(\bm y,\bm{\pi}_{\textrm{int}}^{(i)}, \bm \pi^*) = \bm 0$ \Comment{See \eqref{eq:setup-ift}}
\State  \label{step:solve-sensitivity} Compute $D_{\bm \pi_{\textrm{int}}}h$ and $h(\bm \pi_{\textrm{int}})$ at $\bm{\pi}_{\textrm{int}} = \bm{\pi}_{\textrm{int}}^{(i)}$ 
\State Update $\bm{\pi}_{\textrm{int}}^{(i+1)}$ using  \eqref{eq:outer}, \eqref{eq:NR-outer}, \eqref{eq:Jacobian-outer} and set $i \gets i + 1$
\Until{Termination condition is met} 
\State \textbf{output:} $\bm{\pi}_{\textrm{int}}^{(i+1)}$
\end{algorithmic}
\label{alg:bilevel-NR}
\end{algorithm}
\begin{remark}\label{rem:higher-order-NR}
Since $|N_{\textrm{int}}|$ is typically small, it is possible to contemplate variants of Newton's method with better convergence \cite{third-order-NR,W4} to solve the central system \eqref{eq:outer}.
\end{remark}
\begin{remark} \label{rem:H-solve-interpretation}
The system of equations $H(\bm y,\bm{\pi}_{\textrm{int}}^{(i)}, \bm \pi^*) = \bm 0$ solved at the $i$\textsuperscript{th} iteration in Step \ref{step:solve-H} of Algorithm \ref{alg:bilevel-NR} can be thought of as the network flow equations in \eqref{eq:NF-pi} for the graph $G$ with slack nodes $N_s \cup N_{\textrm{int}}$ and non-slack nodes $N_{\textrm{dep}}$ with potentials $\bm \pi^{(i)}_{\textrm{int}}$ for the additional slack nodes $N_{\textrm{int}}$.
\end{remark}
At this juncture, it would appear from Remark~\ref{rem:H-solve-interpretation} that the reformulation has exacerbated the problem, since we have exchanged the task of solving a large nonlinear system for that of solving a bi-level nonlinear system where a modified nonlinear system of the same size as the original system is required to be solved \emph{in every iteration} (see Step \ref{step:solve-H} of Algorithm \ref{alg:bilevel-NR}).
However, if the Jacobian $D_{\bm y} H$ were block-diagonal, then the solution of $H (\bm y , \bm \pi_{\textrm{int}}^{(i)}, \bm \pi^*) = \bm 0$ would correspond to the solution of  decoupled, smaller nonlinear subsystems corresponding to each of the blocks.
If the network topology could be exploited to engineer the choice of the interface nodes to obtain such decoupled,  smaller subsystems, then it confers a major computational advantage on Algorithm \ref{alg:bilevel-NR}.
Before proceeding on that front, we illustrate the connection our bi-level problem has to the Schur complement \cite{hornjohnson}.

\subsection{Relation to Schur complement}
A Newton-Raphson step for the full system  \eqref{eq:NF-matrix} will lead to $\Delta \bm \pi_s = \bm 0$. Hence, we substitute 
$\bm \pi_s = \bm \pi^*$  and simplify the Newton step to
\begin{gather}
    \begin{bmatrix}
        -D_{\bm f} \bm g^* & \bm B_{\textrm{dep}}^T & \bm B_{\textrm{int}}^T \\ 
        \bm A_{\textrm{dep}} & \bm 0 & \bm 0 \\ 
        \bm A_{\textrm{int}} & \bm 0 & \bm 0
    \end{bmatrix}\;
    \begin{bmatrix} 
        \Delta \bm f \\ \Delta \bm \pi_{\textrm{dep}} \\ \Delta \bm \pi_{\textrm{int}} 
    \end{bmatrix} = 
    - \begin{bmatrix} 
        h_{\mathrm{edge}}(\bm f, \bm \pi) \\ h_{\mathrm{dep}}(\bm f) \\ h_{\mathrm{int}}(\bm f) 
    \end{bmatrix}. \label{eq:full-NR}
\end{gather}
We rewrite \eqref{eq:full-NR} with respect to $\bm y$ defined in \eqref{eq:setup-ift} as
\begin{gather}
    \begin{bmatrix}
        D_{\bm y} H & D_{\bm \pi_{\textrm{int}}} H \\ 
        \bm A_{\textrm{int}} \mathbb{P} & \bm 0
    \end{bmatrix} \;
    \begin{bmatrix}
        \Delta \bm y \\ \Delta \bm \pi_{\textrm{int}}
    \end{bmatrix} = 
    -\begin{bmatrix}
    H\left (\bm y , \bm \pi_{\textrm{int}}, \bm \pi^* \right) \\ h(\bm \pi_{\textrm{int}}) 
    \end{bmatrix}.
    \label{eq:full-NR-y}
\end{gather}
If we eliminate  $\Delta \bm y$, the eliminant can be written using the Schur complement of $D_{\bm y} H$ as
\begin{equation}
\begin{gathered}
     - \bm A_{\textrm{int}}  \mathbb{P} \left[D_{\bm y} H\right]^{-1} \left[ D_{\bm \pi_{\textrm{int}}} H \right]  \Delta \bm \pi_{\textrm{int}}      \\ 
     + \; h(\bm \pi_{\textrm{int}}) \; - \bm A_{\textrm{int}}  \mathbb{P} \left[D_{\bm y} H\right]^{-1} H\left (\bm y , \bm \pi_{\textrm{int}}, \bm \pi^* \right) = \bm 0. 
\end{gathered} \label{eq:schur}
\end{equation}
We know from \eqref{eq:Jacobian-outer} that the iterations of the bi-level problem satisfy $H\left (\bm y , \bm \pi_{\textrm{int}}, \bm \pi^* \right) = \bm 0$ at every Newton step. Hence, \eqref{eq:schur} simplifies and becomes identical to equation \eqref{eq:NR-outer} for $ \Delta \bm \pi_{\textrm{int}}$ .
Thus, the Newton step \eqref{eq:NR-outer} for the bi-level problem \eqref{eq:outer} is equivalent to that induced by the  Schur complement of the Newton step \eqref{eq:full-NR} for the full problem \eqref{eq:NF-matrix}.
\section{Network partitioning with the bi-level formulation}
\label{sec:GNP}
The method of solution with a bi-level formulation has already been explained in detail in the previous section, so our aim is to illustrate how  a judicious choice of interface nodes can result in much smaller subsystems in the bi-level formulation.
Before turning to a  mathematical construction that demonstrates our point,  we first present  an intuitive description.  
\subsection{Intuitive description}
\label{sec:intuitive-description}
\begin{figure}
    \centering
    \includegraphics[width=0.80\linewidth]{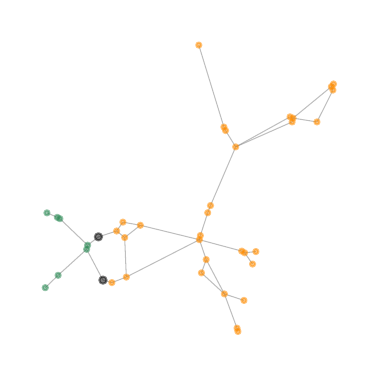}
    \caption{The GasLib-40 network is shown with interface nodes coloured black, while $S_1, S_2$ are shown in different colors with the interface nodes belonging to both. With the interface nodes as slack, can solve two subsystems corresponding to $S_1$ and $S_2$ respectively.}
    \label{fig:partition1}
\end{figure}

Consider the solution of \eqref{eq:NF-pi} specialized for flow of natural gas with the topology implied by GasLib-40 \cite{Schmidt2017} whose graph is depicted in Figure~\ref{fig:partition1}.
From the visual depiction in Figure~\ref{fig:partition1}, it is clear that the two nodes colored in black (interface nodes) allow us to identify two sub-networks (designated $S_1, S_2$) that contain the interface nodes in common.

The solution of the bi-level formulation (Definition~\ref{def:bi-level}) involves the solution of \eqref{eq:NF-pi} for the full network with the interface nodes now designated as additional slack nodes.
However, identification of $S_1, S_2$ allows us to instead solve the network flow systems $\mathcal N \mathcal F(S_1), \; \mathcal N \mathcal F(S_2)$.
These observations in turn raise several questions: 
\begin{enumerate}[label=(Q\arabic*)]
\item \label{q1} How to identify interface nodes for large networks? 
\item \label{q2} Are there any restrictions on the subnetworks? 
\item \label{q3} Why do we expect that  this method will be successful? 
\end{enumerate}
The mathematical development described in the next section will answer the first two questions, but assuming we identify the interface nodes appropriately, we expect success in \ref{q3} since a solution to the system of network flow equations is already known to exist \cite{ss-soln-existence}.

\subsection{Mathematical description}
\label{sec:math-description}
Once again, we denote the graph of the network by $G$, and the edge and vertex sets by $E(G)$ and $V(G)$, respectively. We denote the number of vertices and edges by $|V(G)|$ and $|E(G)|$ respectively.
Note that the removal of a set of vertices (the set can be a singleton) implies the removal of all edges incident on that set of vertices.
We define the following preliminary notions for subsequent use.
%
\begin{definition}[Vertex Separator]
\label{def:vertex-separator}
 A \emph{vertex separator} $\mathcal{C}_V \subset V(G)$ is a non-empty set of vertices whose removal increases the number of connected components. A vertex separator is also known as a \emph{vertex cut} of a graph.
\end{definition}
{If a vertex separator is a singleton, it is called a \emph{cut-point} or \emph{articulation point}. 
However, while a \emph{biconnected} graph has no articulation point, pipeline networks are always planar, non-complete graphs for which vertex separators always exist.
Efficient heuristics \cite{Kernighan1970Feb,Evrendilek2008Feb,Liu1989Sep,Souza2005Jul} and readily available software tools such as METIS \cite{metis, Karypis1997} can easily construct (non-minimal)  vertex separators of various sizes. } 

{The edges between the vertices of the vertex separator are denoted by the set
\begin{equation}
\label{eq:separator-edges}
    E_{\mathcal{C}_V} = \{(v_m, v_n) \in E(G)\; | \; v_m, v_n \in \mathcal{C}_V\}
\end{equation}
}

\begin{definition}[Permissible vertex separator]
\label{def:permissible}
 For this article, a vertex separator $\mathcal{C}_V$ of graph $G$ is  \emph{permissible} if $E_{\mathcal{C}_V} = \emptyset$.
\end{definition}
{From our tests on example instances (GasLib library \cite{Schmidt2017}, the Texas network \cite{birchfield2024structural} etc.), we verified that the criterion $E_{\mathcal{C}_V} = \emptyset$ is not at all restrictive.  
Thus, one can always expect to be able to determine a permissible $\mathcal{C}_V$ with $|\mathcal{C}_V| \ll V(G)$.}
%
%
\begin{definition}[Vertex removal]
 We use the notation $G-\mathcal{C}_V$ to indicate the (possibly disconnected) graph constructed by removal of the vertices in $\mathcal{C}_V \subset V(G)$.   
\end{definition}
We will now formalize the construction of subnetworks associated with a vertex separator in
\begin{lemma}
\label{lemma:subnetworks}
For a connected graph $G$, a 
vertex separator $\mathcal{C}_V \subset V(G)$ can be used to construct $N(\mathcal{C}_V) \geqslant 2$ subnetworks of $G$, $S_1, S_2, \dotsc S_{N(\mathcal{C}_V)}$ such that
\begin{subequations}
    \begin{gather}
    \underset{1 \leqslant k \leqslant N(\mathcal{C}_V)}{\bigcup} E(S_k)  = E(G)\setminus E_{\mathcal{C}_V}, \\
     V(S_i) \cap V(S_j) \subseteq \mathcal{C}_V \quad \forall  \; i\neq j, \; 1 \leqslant i, j \leqslant N(\mathcal{C}_V)
    \end{gather}
    \label{eq:subnetworks}
\end{subequations}
\end{lemma}
 \vspace{-2em}
\begin{proof}
Since $\mathcal{C}_V$ is a vertex separator, $G-\mathcal{C}_V$ will have at least two connected components. The total number will be finite, dependent on the graph as well as the chosen vertex separator. Let us designate this number as $N(\mathcal{C}_V)$, and the connected components as 
$S_1^-, S_2^-, \dotsc S_{N(\mathcal{C}_V)}^-$.
By their very definition, it is clear that
\begin{subequations}
    \begin{gather}
    \underset{1 \leqslant k \leqslant N(\mathcal{C}_V)}{\bigcup} E(S_k^-)  = E(G-\mathcal{C}_V), \\
     V(S_i^-) \cap V(S_j^-) = \emptyset \quad \forall  \; i\neq j, \; 1 \leqslant i, j \leqslant N(\mathcal{C}_V)
    \end{gather}
    \label{eq:connected-comp}
\end{subequations}
The subnetworks with the desired properties follow via the following construction that adds selected edges to the connected components $S_1^-, S_2^-, \dotsc S_{N(\mathcal{C}_V)}^-$ as follows: for each $v_1 \in \mathcal{C}_V$, if there is any edge in $E(G)$ that connects it to some  $v_2 \in V(S_k^-)$, then $S_k^-$ is augmented by adding the edge $(v_1, v_2)$. Thus,  all edges in $E(G)$ would be added except for those in $E_{C_V}$, and we now have  $S_1, S_2, \dotsc S_{N(\mathcal{C}_V)}$ whose vertices and edges satisfy the stated properties in the lemma.
\end{proof}
If the vertex separator be permissible as in Definition~\ref{def:permissible}, then we obtain subnetworks like that  illustrated in Figure~\ref{fig:partition1}. 

The construction of the subnetworks corresponding to a permissible vertex separator subtly modifies
our proposed method for solving the bi-level formulation \eqref{eq:outer} due to  
{
\begin{theorem}
\label{thm}
    For a network $G$ with a permissible vertex separator $\mathcal{C}_V$, the network flow system $H(\bm y,\bm{\pi}_{\textrm{int}}^{(i)}, \bm \pi^*) = \bm 0$ has a Jacobian $D_{\bm y} H$ that is block-diagonal.
\end{theorem}
}
%
 \vspace{-1em}
 {
\begin{proof}
The vertices in $\mathcal{C}_V$ serve the same role as $N_{\textrm{int}}$ defined in Section~\ref{sec:bi-level}, hence we assume that $\mathcal{C}_V \cap V_s = \emptyset$. 
According to Lemma~\ref{lemma:subnetworks}, the permissible vertex separator $\mathcal{C}_V$ partitions the graph $G$ into 
$N(\mathcal{C}_V)$ subnetworks, denoted $S_1, S_2, \dotsc, S_{N(\mathcal{C}_V)}$. 
Thus one can partition  $\bm y$ as  $[\bm y_1, \bm y_2, \dotsc \bm y_{N(\mathcal{C}_V)}]^T$, and (with a slight abuse of notation)
$H(\bm y,\bm{\pi}_{\textrm{int}}^{(i)}, \bm \pi^*)$  as
$[H_1, H_2, \dotsc, H_{N(\mathcal{C}_V)}]^T$ with
$H_k = H(\bm y_k,\bm{\pi}_{\textrm{int}}^{(i)}, \bm \pi^*)$.
Thus the system  $H(\bm y,\bm{\pi}_{\textrm{int}}^{(i)}, \bm \pi^*)$ or $\mathcal{N}\mathcal{F}(G)$  is decomposed into $N(\mathcal{C}_V)$ independent subnetwork systems $H(\bm y_k,\bm{\pi}_{\textrm{int}}^{(i)}, \bm \pi^*)$ corresponding to the subnetwork $S_k$, consistent with~\eqref{eq:subnetworks}. 
Since $E_{\mathcal{C}_V} =\emptyset$ for a permissible vertex separator, the systems are decoupled and  the Jacobian is block-diagonal.
\end{proof}
}

{
In brief, Theorem~\ref{thm} has traded the complexity of iterative solution of a large nonlinear system  $\mathcal N \mathcal F(G)$ (see Remark~\ref{rem:H-solve-interpretation}) for the iterative solution of a collection of decoupled, smaller nonlinear subsystems $\mathcal N \mathcal F(S_k)$ for each $1\leqslant k \leqslant N(\mathcal{C}_V)$ to gain computational efficiency in Step~\ref{step:solve-H} of Algorithm~\ref{alg:bilevel-NR}.
Similarly, in Step~\ref{step:solve-sensitivity}, the sensitivity or Jacobian contribution is also assembled from each subnetwork as follows: For the interface nodes in $\mathcal{C}_V \cap V(S_k)$, the sensitivity of the solution with respect to the corresponding components of $\bm \pi_{\textrm{int}}$ is calculated as stated in Remark~\ref{rem:sensitivity}. 
However, for interface nodes in $\mathcal{C}_V \setminus V(S_k)$, the corresponding sensitivities are zero.
}

%
\begin{remark}
\label{rem:2-node-submetwork}
  It is necessary to have $|V(S_k)| > 2$. This is because, if $|V(S_k)| = 2$, both vertices are interface nodes with potentials known and the edge flow then becomes indeterminate for compressors or pressure regulators (where $g^*_{ij} =0$). 
\end{remark}
%
%
%
We now have answers to the questions \ref{q1} and \ref{q2} that were posed at the conclusion of section~\ref{sec:intuitive-description}.
It is clear that there are graph-theoretic algorithms and software that can be used to identify vertex separators for use in this problem, answering \ref{q1} while Remark~\ref{rem:2-node-submetwork} addresses \ref{q2}.
\subsection{Observations and Comments}


The underlying philosophy of the method we propose (to partition and solve the system)  is identical to the hierarchical network partitioning (HNP) method \cite{network-partitioning}. 
The  vertex separator in HNP consists solely of articulation points, each of which are themselves vertex separators.  
HNP proceeds with an initial flow solve after which each subnetwork is solved once, with the partitions traversed hierarchically starting from the partition that contains all the slack nodes.  
The initial flow solve in HNP  allows the determination of  flow in two node subnetworks unlike in GNP as explained in Remark~\ref{rem:2-node-submetwork}.
We now detail the major advantages of our proposed method over HNP. 

%
%
\begin{figure}
    \centering
    \includegraphics[width=0.90\linewidth]{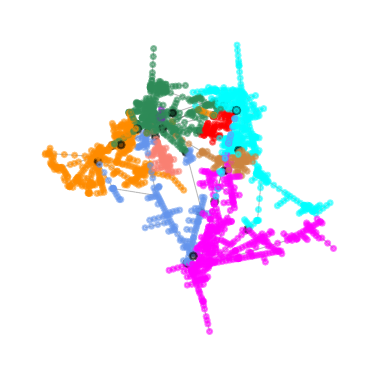}
    \caption{Partitioning of Texas network with 2451 vertices \cite{birchfield2024structural} into 9 subnetworks, all of size less than 500 vertices. The 25 interface nodes are coloured black.}
    \label{fig:texas}
\end{figure}
\begin{figure}
    \centering
    \includegraphics[scale=0.6]{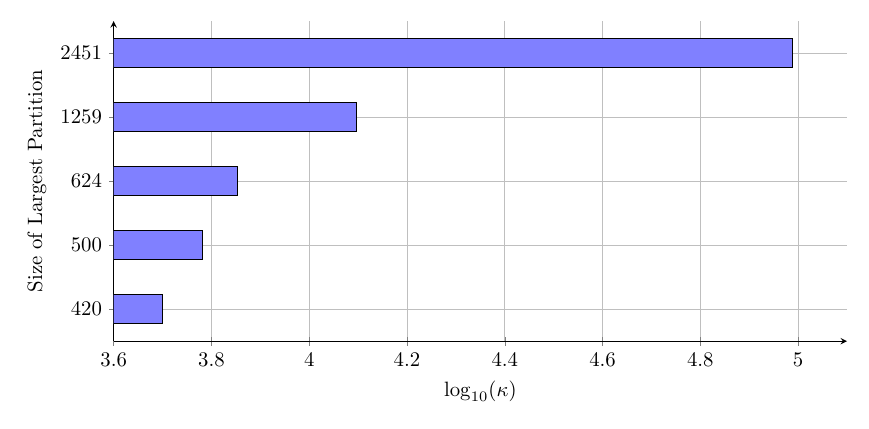}
    \caption{The bar graph shows the logarithm of the maximum condition number of the Jacobian, $\kappa$, corresponding to partitions of different sizes. The Texas network has 2451 vertices, and we see how partitioning the network reduces the condition number and makes the system easier to solve.}
    \label{fig:cond}
\end{figure}

\noindent \textbf{Balanced Partitions}: 
Since the vertex separator in HNP consists solely of articulation points, cycles (biconnected subnetworks) cannot be broken up further. Thus, for the Texas network \cite{birchfield2024structural} with 2451 vertices, all partitions with HNP will have a subnetwork that contains \emph{at least} 882 vertices.  However, GNP allows balanced partitions so that by means of 25 interface nodes (see Fig~\ref{fig:texas}), the Texas network is partitioned into 9 subnetworks, all of which have \emph{less than} 500 vertices. 

\noindent  \textbf{Positioning of Slacks}:
If there are multiple slack nodes, HNP requires that they belong to the \emph{same} subnetwork, which renders HNP inapplicable or useless when the slack nodes are far apart geographically.
In contrast, GNP does not limit the number or positions of slack nodes within the partitions. 

\noindent \textbf{Multinetwork Context}:
Various regional and local networks in the US are tied through multiple interconnects (to ensure redundancy) as mentioned earlier in Section~\ref{sec:intro}.  
 These interconnects are natural vertex separators (Definition~\ref{def:vertex-separator}) but not articulation points. Thus, while HNP is inapplicable, GNP is a viable technique that limits data sharing solely to the interconnects between networks, leaving operators with leeway to use their own legacy solvers or software within their domain. 

\noindent \textbf{Generalizability}: 
HNP is predicated on the  fact that for any edge, if the flow has a certain magnitude and sign relative to one end, then it must have the same magnitude but a reversed sign relative to the other end. However, this property, termed ``lossless flow", is not a requirement for the applicability of GNP.  While beyond the scope of this article,  if we consider AC power flow with \emph{lossy} transmission \cite{coffrin2016acopf}, HNP is clearly inapplicable, but GNP could be used to solve the system of AC power flow equations.  

\subsection{Results}
Given the fundamental premise of this article to solve the network flow system via partitioned subsystems,  proof-of-concept  experiments that demonstrate and verify the viability of the method  are of most interest.
Accordingly, the approach  was tested on the GasLib networks \cite{Schmidt2017} with 11, 24, 40, 134 and 582 vertices as well as the Texas network (Figure~\ref{fig:texas}), and in every case, varying the number and placement of the slack nodes, starting with partitions of various sizes, the method was successful. 
Moreover, the condition number of the Jacobian decreased with the size of the partitions in every instance (shown here for Texas network in Figure~\ref{fig:cond}), further supporting the underlying motivation for developing the method. 

\section{Conclusion}
We presented an algorithm  that utilizes a given partition of the 
network into tractable sizes to compute
a global solution for the full nonlinear system through local
solution of smaller subsystems induced by the partitions. 
If the partitions are induced by interconnects or transfer points corresponding to networks owned by different operators, the method ensures data sharing occurs solely at the interconnects, leaving  network operators free to  solve the network flow system corresponding to  their own domain in any manner of their choosing. 
The connection between the proposed method and the Schur complement was clarified  and the  method's viability demonstrated on some challenging test cases.
Future work could look to adapt this method to the solution of the single-phase AC power flow equations.



\bibliographystyle{IEEEtran}
\bibliography{ref}

\end{document}